\documentclass[aps,pra,twocolumn,footinbib,superscriptaddress]{revtex4-1}
\usepackage{graphicx}
\usepackage{indentfirst}
\usepackage{braket}
\usepackage{float}
\usepackage{amsmath}
\usepackage{epstopdf}
\usepackage{CJK}
\usepackage{esint}
\usepackage{color}
\usepackage[T1]{fontenc}
\usepackage{subfigure}
\usepackage{letltxmacro}
\LetLtxMacro{\ORIGselectlanguage}{\selectlanguage}
\makeatletter
\DeclareRobustCommand{\selectlanguage}[1]{%
  \@ifundefined{alias@\string#1}
    {\ORIGselectlanguage{#1}}
    {\begingroup\edef\x{\endgroup
       \noexpand\ORIGselectlanguage{\@nameuse{alias@#1}}}\x}%
}
\newcommand{\definelanguagealias}[2]{%
  \@namedef{alias@#1}{#2}%
}
\makeatother

\definelanguagealias{en}{english}
\usepackage{amsfonts}
\usepackage{footmisc}
\usepackage{scrextend}
\usepackage{multirow}
\usepackage{cleveref}
\usepackage[english]{babel}
\usepackage{url}

\newcommand{\qed}{\nobreak \ifvmode \relax \else
      \ifdim\lastskip<1.5em \hskip-\lastskip
      \hskip1.5em plus0em minus0.5em \fi \nobreak
      \vrule height0.75em width0.5em depth0.25em\fi}
      
\def\be{\begin{equation}}
\def\ee{\end{equation}}
\def\ba{\begin{eqnarray}}
\def\ea{\end{eqnarray}}

\begin{document}

\date{\today}

\title{High-order encoding schemes for floodlight quantum key distribution}
\author{Quntao Zhuang}
\email{quntao@mit.edu}
\affiliation{Department of Physics, Massachusetts Institute of Technology, Cambridge, Massachusetts 02139, USA}
\affiliation{Research Laboratory of Electronics, Massachusetts Institute of Technology, Cambridge, Massachusetts 02139, USA}
\author{Zheshen Zhang}
\affiliation{Research Laboratory of Electronics, Massachusetts Institute of Technology, Cambridge, Massachusetts 02139, USA}
\affiliation{Present address:  Department of Materials Science and Engineering, the University of Arizona, Tucson, Arizona 85721, USA}
\author{Jeffrey H. Shapiro}
\affiliation{Research Laboratory of Electronics, Massachusetts Institute of Technology, Cambridge, Massachusetts 02139, USA}%

\begin{abstract} 
Floodlight quantum key distribution (FL-QKD) has realized a 1.3\,Gbit/s secret-key rate (SKR) over a 10-dB-loss channel against a frequency-domain collective attack [Quantum~Sci.~Technol. {\bf 3}, 025007 (2018)].  It achieved this remarkable SKR by means of binary phase-shift keying (BPSK) of multiple optical modes.  Moreover, it did so with available technology, and without space-division or wavelength-division multiplexing.  In this paper we explore whether replacing FL-QKD's BPSK modulation with a high-order encoding can further increase that protocol's SKR.   First, we show that going to $K$-ary phase-shift keying with $K = 32$ doubles---from 2.0 to 4.5\,Gbit/s---the theoretical prediction from [Phys.~Rev.~A {\bf 94}, 012322 (2016)] for FL-QKD's BPSK SKR on a 50-km-long fiber link.  Second, we show that $2d\times 2d$ quadrature amplitude modulation (QAM) does not offer any SKR improvement beyond what its $d=1$ case---which is equivalent to quadrature phase-shift keying---provides. 
\end{abstract} 

\maketitle

\section{Introduction}

Quantum key distribution~\cite{Bennett2014} (QKD) allows remote parties (Alice and Bob) to create a shared random bit string with unconditional security.  Later, they can employ their shared string for one-time-pad (OTP) encryption~\cite{Shannon1949} of messages they wish to keep entirely private from any eavesdropper (Eve).  Unfortunately, current QKD systems' Mbit/s secret-key rates (SKRs) \cite{Lucamarini2013,Huang2015,Zhong2015,Lee2016,Islam2017} fall far short of what is needed to make high-speed (Gbit/s) transmission with OTP encryption ready for widespread deployment.  Floodlight QKD (FL-QKD)~\cite{Zhuang2016,Zhang2017,Zhang2018} is a recent protocol that uses binary phase-shift keying (BPSK) of multiple optical modes and homodyne detection to achieve security against the optimum frequency-domain collective attack.  Its initial theoretical study~\cite{Zhuang2016} predicted that FL-QKD was capable of Gbit/s SKRs at metropolitan-area distances over single-mode fiber (no space-division multiplexing) in a single-wavelength channel (no wavelength-division multiplexing) without the need to develop any new technology.   The initial table-top experimental demonstration of FL-QKD~\cite{Zhang2017} used 100\,Mbit/s modulation to realize a 55\,Mbit/s SKR over a 10-dB-loss channel (equivalent to 50\,km of low-loss fiber) in a setup that was limited by the bandwidth of its electronics.  A subsequent table-top experiment~\cite{Zhang2018}, using GHz-bandwidth electronics, attained a 1.3\,Gbit/s SKR over a 10-dB-loss channel using a 7\,Gbit/s modulation rate  

Why is FL-QKD's SKR so much higher than prior state of the art, even when compared at the same collective-attack security level?  It is because FL-QKD takes advantage of multimode encoding, whereas the predominant decoy-state BB84 protocol does not~\cite{Lucamarini2013}, and conventional continuous-variable (CV) QKD protocols require single-mode encoding~\cite{Huang2015}.  Moreover, the SKR advantage over decoy-state BB84 shown in recent $D$-dimensional QKD experiments~\cite{Zhong2015,Lee2016,Islam2017} comes only from mitigating the SKR-limiting effect of single-photon detectors' dead time, i.e., within each time slot of their $D$-slot symbols these protocols take no advantage of multimode encoding.  Thus the SKRs in \emph{bits/s} for prior state-of-the-art systems are constrained to be no more than the ultimate limit on SKR in \emph{bits/mode}, viz., the PLOB bound~\cite{Pirandola2017}, ${\rm SKR} \le -\log_2(1-\eta)$\,bits/mode for a channel with transmissivity $\eta$.  In contrast, FL-QKD's predicted 2\,Gbit/s SKR~\cite{Zhuang2016} over a 50-km-long fiber link uses 10\,Gbit/s BPSK modulation of 200\,modes/symbol making its $10^{-3}$\,bits/mode well below the $-\log_2(1-\eta) = 0.15$ PLOB bound for $\eta = 0.1$, while still affording the Gbit/s SKRs needed for high-speed OTP encryption.  

How can we increase FL-QKD's bits/mode SKR, other things being equal, to further enhance its bits/sec SKR?  Because FL-QKD relies on homodyne detection, there is a potential answer from classical fiber-optic communication, where a similar problem has been confronted in the context of increasing the spectral efficiency (bits/s-Hz = bits/mode) for coherent (homodyne or heterodyne) detection systems~\cite{Essiambre2012}.  For classical communication the answer is to go to a high-order modulation format, e.g., $K$-ary phase-shift keying (KPSK) or quadrature amplitude modulation (QAM).  Therefore, in this paper we will evaluate the merits of FL-QKD's using those formats.  We show that KPSK with $K = 32$ doubles---from 2.0 to 4.5\,Gbit/s---the theoretical prediction from Ref.~\cite{Zhuang2016} for FL-QKD's BPSK SKR on a 50-km-long fiber link, but we find that $2d\times 2d$-symbol QAM does not offer any SKR improvement beyond what its $d=1$ case---which is equivalent to quadrature phase-shift keying, i.e., 4PSK---provides. 

The remainder of the paper is organized as follows.  We begin, in Sec.~\ref{Protocol}, by extending the FL-QKD protocol---as presented in Ref.~\cite{Zhuang2016} and subsequently realized in Refs.~\cite{Zhang2017,Zhang2018}---to allow for high-order encoding, using either the KPSK or QAM signal constellations.  Next, in Sec.~\ref{SKRs}, we analyze FL-QKD's performance when it employs KPSK with $1\le \log_2(K) \le 5$, or $2d\times 2d$ square-lattice QAM with $1\le d \le 4$.  We conclude, in Sec.~\ref{Discussion}, with some discussion and suggestions for future work.  Derivation details appear in Appendices~\ref{AppPSK} and \ref{AppQAM}.

\section{FL-QKD with High-Order Encoding\label{Protocol}}
\begin{figure}[h]
\includegraphics[width=0.45\textwidth]{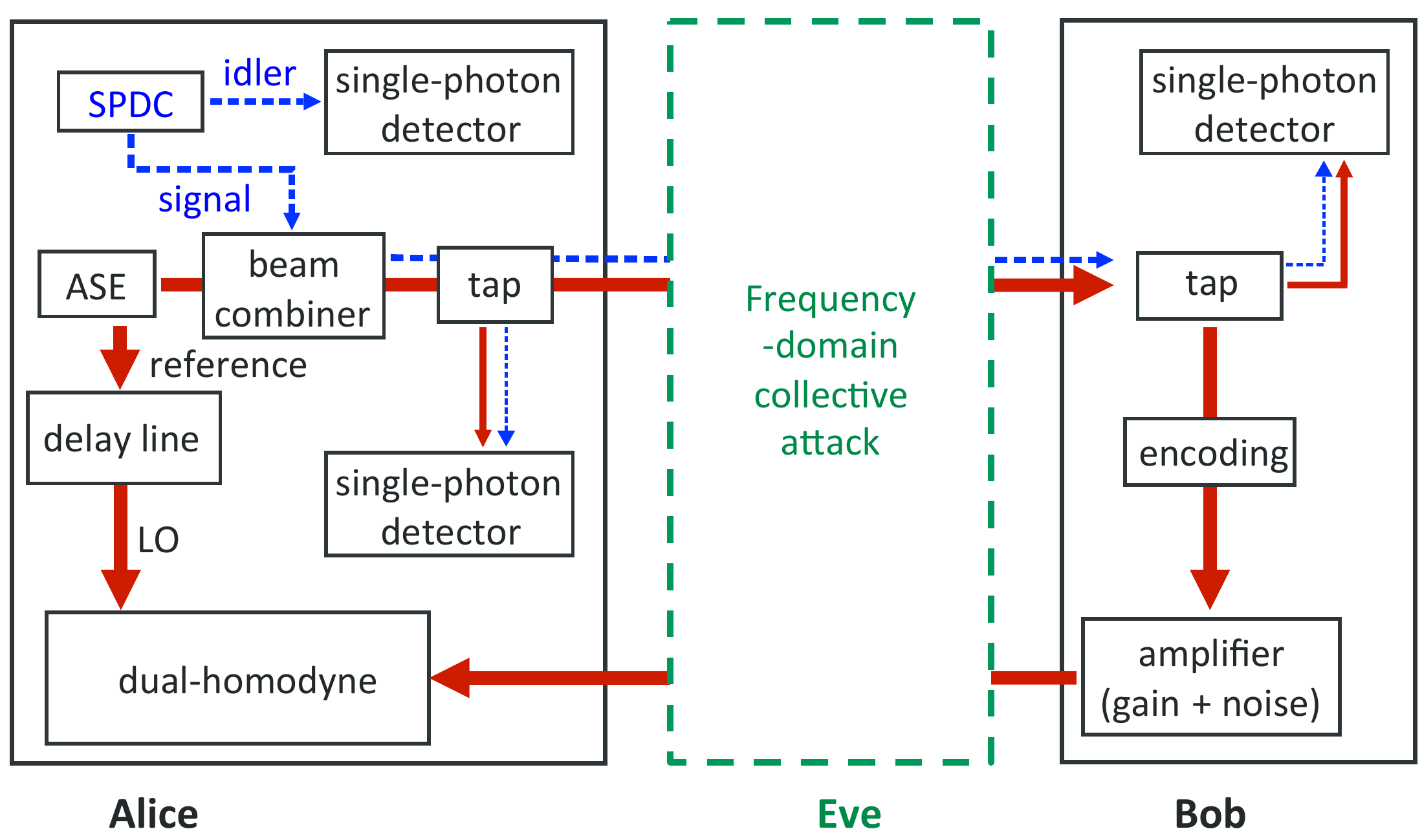}
\caption{Quantum channel setup for FL-QKD under frequency-domain collective attack.  ASE:  amplified spontaneous emission source.  SPDC:  spontaneous parametric downconverter. LO:  local oscillator.}
\label{scheme_FLQKD}
\end{figure}

In FL-QKD with high-order encoding~(schematic in Fig.~\ref{scheme_FLQKD}), Alice splits the $W$-Hz bandwidth, flat-top spectrum, high-brightness~(many photons/mode) output from an amplified spontaneous emission~(ASE) source into a low-brightness~($\ll 1$ photon/mode) signal and a high-brightness reference. To enable channel monitoring, Alice combines her low-brightness ASE with the signal output from a spontaneous parametric downconverter (SPDC)---of the same $W$-Hz bandwidth flat-top spectrum as the ASE---in an $n$:1 ASE-to-SPDC-ratio with $n \gg 1$.  Alice uses a single-photon detector to monitor her SPDC's idler and another single-photon detector to monitor a $\kappa_A \ll 1$ fraction that she taps from her combined ASE-SPDC light, while sending the remainder of that light---whose brightness is $N_S \ll 1$ photon/mode---to Bob.  Alice retains her bright reference beam in an optical-fiber delay line---using amplifiers as needed---for use as her dual-homodyne receiver's high-brightness ($N_{\rm LO} \gg 1$ photons/mode) local oscillator (LO).  

In the absence of Eve, the fiber link from Alice to Bob is a pure-loss channel with transmissivity $\kappa_S\ll1$. Eve's presence, however, allows her to control that channel, hence Alice and Bob must perform channel monitoring to bound Eve's information gain.  So, prior to his encoding operation, Bob taps a small fraction $\kappa_B\ll 1$ of the light he receives and sends it to a single-photon detector.  The outputs from Alice and Bob's single-photon detectors enable them to determine the singles rates $S_I$ for Alice's idler and $S_A$ ($S_B$) for Alice's (Bob's) tap, as well as $C_{IA}$ ($C_{IB}$) and $\tilde{C}_{IA}$ ($\tilde{C}_{IB}$) the time-aligned and time-shifted coincidence rates between Alice's idler and Alice's (Bob's) tap.  They use their measurements to:  (1) verify that Bob receives the photon flux he would get were Eve absent; and (2) determine Eve's intrusion parameter, $f_E$, from~\cite{Zhuang2016}
\begin{equation}
f_E = 1 - [(C_{IB}-\tilde{C}_{IB})/S_B]/[(C_{IA}-\tilde{C}_{IA})/S_A].
\end{equation}
Alice and Bob's knowing Eve's intrusion parameter quantifies the integrity of the Alice-to-Bob channel, and allows them to place an upper bound on Eve's Holevo-information rate for her optimum frequency-domain collective attack.  Eve can realize that optimum attack in the form of an SPDC light-injection attack~\cite{Zhuang2016}, in which case $f_E$ is the fraction of the light entering Bob's terminal that comes from Eve.  

\begin{figure}[h]
\subfigure[]{
\includegraphics[width=0.2\textwidth]{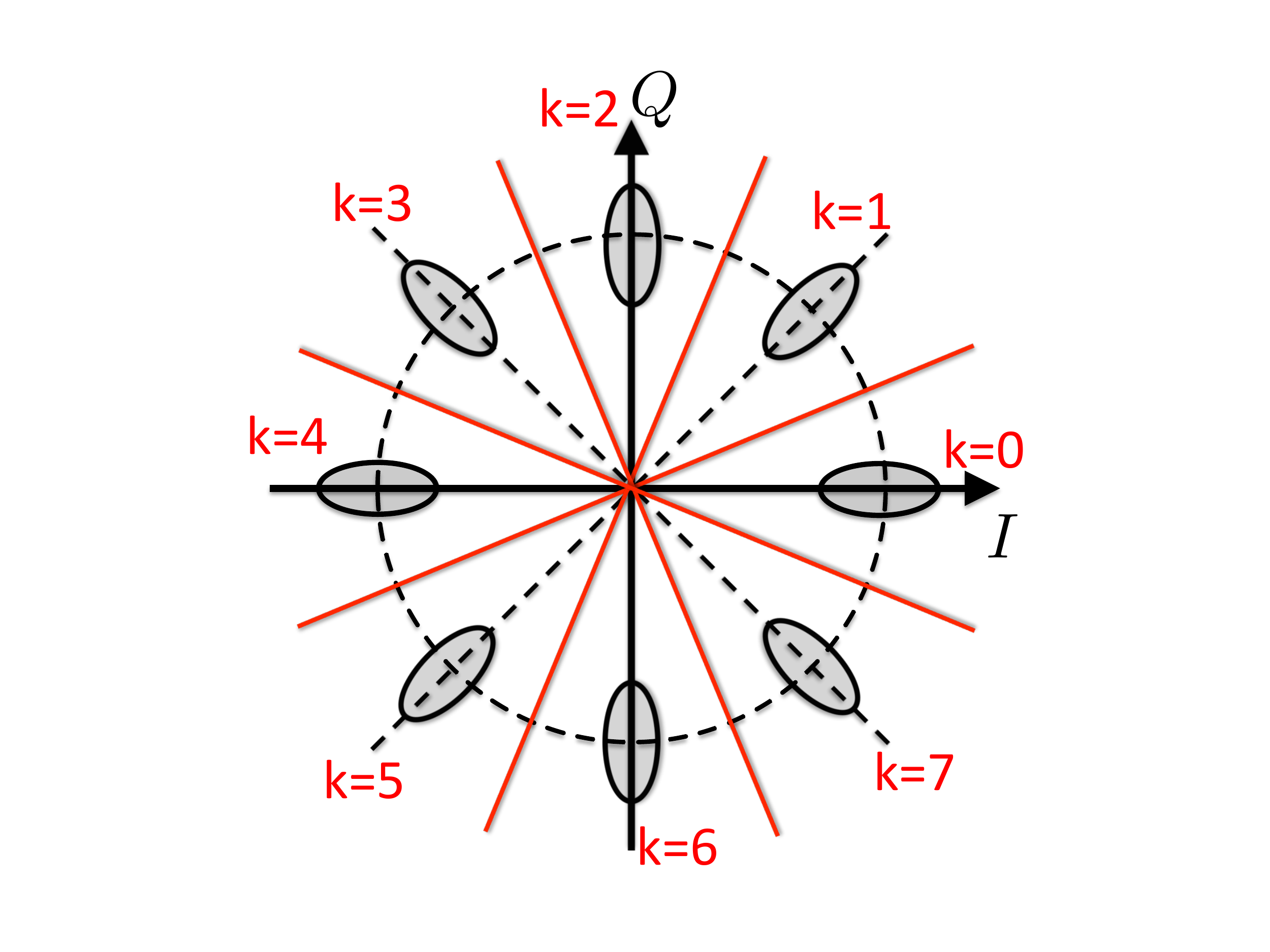}
\label{scheme_modulation}}
\subfigure[]{
\includegraphics[width=0.2\textwidth]{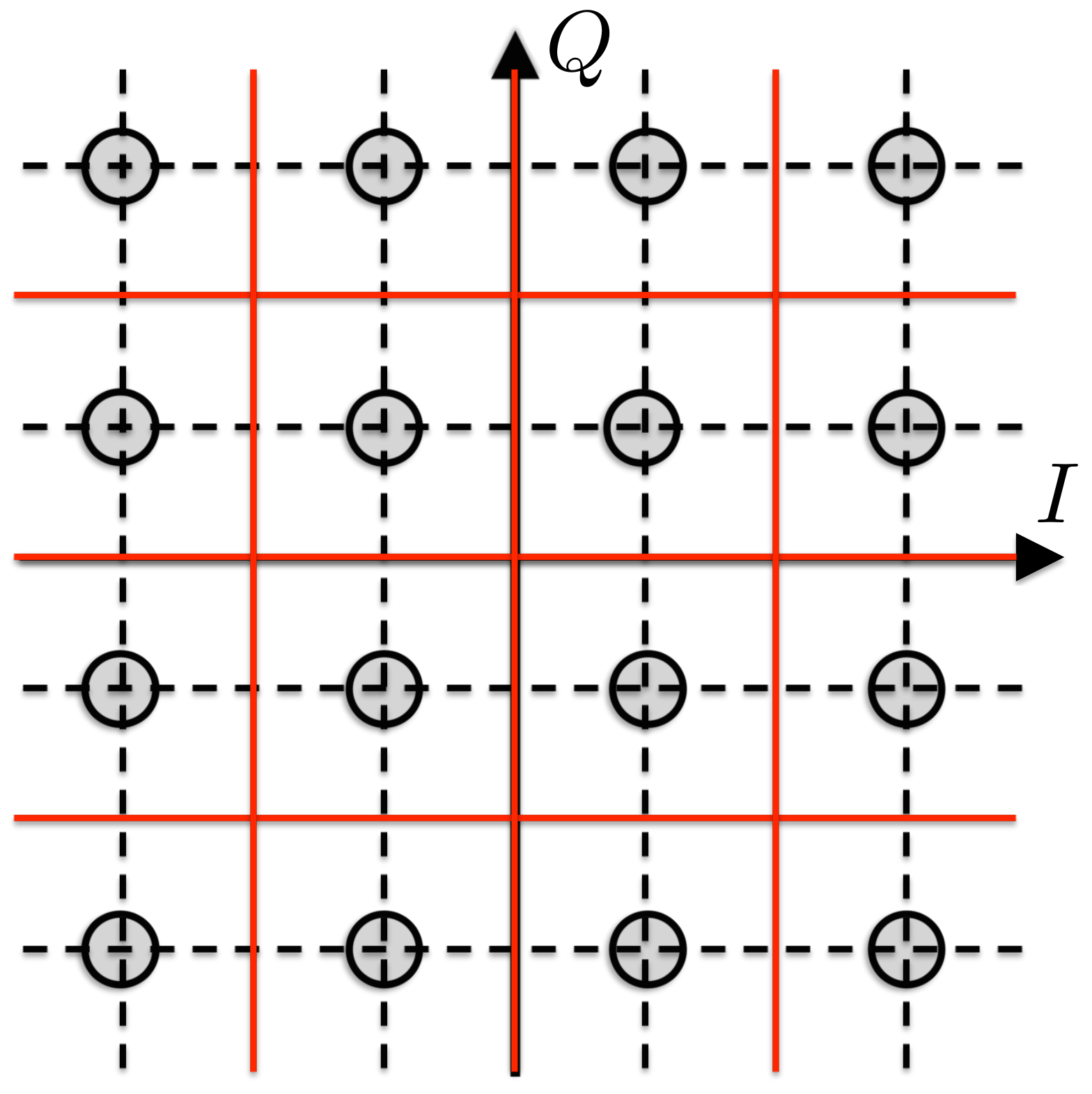}
\label{scheme_modulation_QAM}}
\caption{Signal constellation examples:  (a) 8PSK, (b) $4\times 4$ QAM. In both cases the gray shading marks one-standard-deviation regions for Alice's receiver about the $\{\bar{I}_k+i\bar{Q}_k\}$---where $\bar{I}_k+i\bar{Q}_k$ is how Bob's $k$th transmitted symbol would appear, in a noise-free world, at the output of Alice's dual-homodyne receiver---and the red lines mark the boundaries of her minimum error-probability decision regions.}
\end{figure}

To complete his part of the FL-QKD protocol, Bob first takes the light not routed to his channel monitor's single-photon detector and modulates it with a random symbol selected from his signal constellation at an $R = 1/T$\,baud symbol rate.  In Refs.~\cite{Zhuang2016,Zhang2017,Zhang2018}, that constellation was BPSK, i.e., Bob randomly applied a 0\,Rad or $\pi$\,Rad phase shift.  In the present work, Bob will employ either a KPSK or a square-lattice QAM constellation, as shown in Figs.~\ref{scheme_modulation} and \ref{scheme_modulation_QAM}, respectively, and detailed in Sec.~\ref{SKRs}. 
\cite{Shapiro2009}.

After his encoding, Bob amplifies his modulated light with a gain $G_B \gg 1$ quantum-limited amplifier whose output ASE has brightness $N_B=G_B-1$,  and sends the modulated and amplified light back to Alice through what, in Eve's absence, is a $\kappa_S$-transmissivity fiber.  The amplifier's gain will overcome the return-path loss insofar as Alice is concerned,  thus making FL-QKD's performance only subject to one-way path loss, despite its being a two-way protocol.  Furthermore, the amplifier's ASE will mask Bob's modulation from Eve's passive (listening only) attack~

To decode Bob's symbols, Alice uses dual-homodyne reception, i.e., she 50-50 beam splits both the light returned from Bob and her LO, and then makes homodyne measurements of the $I$ (0\,rad LO phase shift) and $Q$ ($\pi/2$\,rad LO phase shift) in-phase and quadrature components of the returned light, as in classical KPSK or QAM fiber-optic communications.  Following Alice's minimum error-probability decoding of Bob's symbol stream from her measured sequence of $I+iQ$ values, Alice and Bob complete the FL-QKD protocol with the usual key reconciliation and privacy amplification steps, using an authenticated classical communication channel.  

\section{Secret key rates\label{SKRs}}

Our route to determining FL-QKD's performance when it employs KPSK or QAM parallels what was done in Ref.~\cite{Zhuang2016} for FL-QKD using BPSK:  we obtain a lower bound on the SKR from 
\begin{equation}
{\rm SKR}_{\rm LB} = \beta I_{AB} - \chi_{EB}^{\rm UB}.
\label{SKRformula}
\end{equation}
Here: (1) $\beta$ is Alice and Bob's reconciliation efficiency; $I_{AB}$ is Alice and Bob's bits/s Shannon-information rate, which they calculate from their measured conditional probability distribution---obtained during reconciliation---for Alice's decoded symbol given Bob's encoded symbol; and (2) $\chi_{EB}^{\rm UB}$ is an upper bound on Eve's bits/s Holevo-information rate, which Alice and Bob calculate from their channel monitors' $f_E$ measurement.  Also, we are assuming that:  (1) Eve mounts an SPDC light-injection attack---which, see below, realizes her optimum frequency-domain collective attack on KPSK---with $f_E = 0.01$ intrusion parameter; and (2) ${\rm SKR}_{\rm LB}$ is sufficiently high that a typical QKD session will push deep into the asymptotic regime, i.e., no finite-key correction is needed.  

The subsections that follow evaluate ${\rm SKR}_{\rm LB}$ for FL-QKD with KPSK and QAM using the same parameter values that Ref.~\cite{Zhuang2016} assumed, thus enabling direct comparisons of FL-QKD's performance using these high-order encodings to the protocol's performance with its original BPSK encoding.  These parameter values are as follows.  (1) Alice's ASE and SPDC sources operate at 1550\,nm wavelength and have $W=2$\,THz bandwidth. (2) Alice's transmission to Bob has a 99:1 ASE-to-SPDC ratio, and its brightness, $N_S$, is chosen, for each propagation distance, to maximize ${\rm SKR}_{\rm LB}$. (3) Alice and Bob are connected by $L$-km-long single-mode fibers with 0.2\,dB/km loss.  (4) Alice and Bob use 1\% taps for their channel monitors. (5) Bob's symbol rate is $R = 10$\,Gbaud ($T = 0.1$\,ns symbol duration)~\cite{footnote1}. (6) Bob's amplifier has gain $G_B = 10^4$. (6) Alice's LO is undegraded with brightness $N_{\rm LO} = 10^4$, and her receiver has an $\eta = 0.9$ homodyne efficiency.  (7) Alice and Bob's reconciliation efficiency is $\beta = 0.94$.  

Before proceeding to our ${\rm SKR}_{\rm LB}$ evaluations, there is an important point to make about Alice's homodyne-measurement statistics.  With our assumed $T = 0.1$\,ns symbol duration and $W=2$\,THz source bandwidth, there are $M = TW = 200\,$modes/symbol in Bob's transmission to Alice.  In this $M\gg 1$ regime, the central limit theorem implies that Alice's $I$ and $Q$ values for each received symbol are jointly Gaussian random variables given the value of Bob's transmitted symbol.  Furthermore, the means, variances, and covariance of $I$ and $Q$---which fully characterize their joint conditional distribution---can be obtained from the value of the transmitted symbol and the conditional covariance matrix of the return and reference beams' $M$ independent identically-distributed (iid) return-LO mode pairs.

\subsection{FL-QKD Performance with KPSK Encoding\label{KPSK}}
In FL-QKD with KSPK encoding, Bob's applies a $2\pi k/K$\,rad phase shift to the light remaining after his monitor tap, where $k$ (his symbol to be encoded) is equally likely to be any integer between 0 and $K-1$.  As a result, given Bob's transmitted symbol $k$, the joint distribution for Alice's $I$ and $Q$ has the rotational symmetry shown in Fig.~\ref{scheme_modulation}.  Specifically, $p(\,I,Q\mid k\,)$ is a Gaussian, whose mean $\langle I + iQ\rangle = \bar{I}_k+i\bar{Q}_k$ has phase angle $2\pi k/K$ and a $k$-independent magnitude, and whose covariance matrix is such that
\begin{align}
\tilde{I}_k &\equiv I\cos(2\pi k/K) + Q\sin(2\pi k/K) \label{tildeI}\\[.05in]
\tilde{Q}_k &\equiv -I\sin(2\pi k/K) + Q\cos(2\pi k/K) \label{tildeQ}
\end{align}
are statistically independent with $k$-independent variances $\sigma^2_{\tilde{I}} > \sigma^2_{\tilde{Q}}$.   (See~Appendix~\ref{AppPSK} for the details.)  

The preceding statistics make it easy to determine Alice's minimum error-probability rule for decoding Bob's transmitted symbol from her $I+iQ$ measurement.  Because Bob sends each possible symbol with equal probability, the minimum error-probability rule reduces to making a maximum-likelihood decision as to which symbol was sent~\cite{Wozencraft1965}.   Because the conditional statistics of $I$ and $Q$ are Gaussian and rotationally symmetric, the maximum-likelihood decision rule is minimum-distance decoding:  Alice decodes her measured $I+iQ$ as the symbol whose $\bar{I}_k+i\bar{Q}_k$ is closest to that measured value.   As shown in Fig.~\ref{scheme_modulation}, this means that the decision region,  $\mathcal{D}_k$, in the $I+iQ$ plane wherein Alice decodes symbol $k$ is, 
\begin{equation}
\mathcal{D}_k = \{\,I+iQ : -\pi k/K \le \theta < \pi k/K\},
\end{equation}
where $|I+iQ|e^{i\theta}$ is the polar-coordinate form of $I+iQ$. 

Using the $\{\mathcal{D}_k\}$, together with the equiprobable nature of Bob's encoding and the jointly-Gaussian conditional distributions $\{p(\,I,Q\mid k\,)\}$, we can numerically evaluate the conditional probabilities $\Pr(\,\tilde{k}\mid k\,)$ for Alice to decode her $I+iQ$ value as $\tilde{k}$, given that Bob sent symbol $k$, via
\begin{equation}
\Pr(\,\tilde{k}\mid k\,) = \int\!\int_{\mathcal{D}_{\tilde{k}}}\!{\rm d}I{\rm d}Q\,p(\,I,Q\mid k\,).
\label{transmatrix}
\end{equation}
Alice and Bob's Shannon-information rate then follows from 
\begin{equation}
I_{AB}=R\!\left\{\sum_{k=0}^{K-1}\sum_{\tilde{k}=0}^{K-1}\frac{\Pr(\,\tilde{k}\mid k\,)}{K}
\log_2\!\left[\frac{K\Pr(\,\tilde{k}\mid k\,)}{\sum_{k'=0}^{K-1}\Pr(\,\tilde{k}\mid k'\,)}\right]\right\}.
\label{IAB_PSK}
\end{equation}

At this point, we can obtain ${\rm SKR}_{\rm LB}$ from Eq.~(\ref{SKRformula}) once we have an upper bound on Eve's Holevo-information rate, $\chi^{\rm UB}_{EB}$, when she mounts an SPDC light-injection attack.  In that attack, Eve injects signal light from her own $W$-Hz bandwidth, flat-top spectrum, low-brightness SPDC source into the Alice-to-Bob channel, so that it will get modulated and amplified by Bob and then transmitted on the Bob-to-Alice channel.  Eve also stores her SPDC source's idler light, for use as a reference and then---making use of that reference, plus the light she has tapped from the Alice-to-Bob channel, and the light she taps from the Bob-to-Alice channel---Eve makes the collective quantum measurement that maximizes her Holevo-information rate.  

In Ref.~\cite{Zhuang2016} it was shown that Eve's SPDC light-injection attack realizes her optimum frequency-domain collective attack when Bob uses BPSK encoding.  It turns out that this is still true when Bob uses KPSK encoding,  because KPSK's signal constellation is rotationally symmetric.  This makes Eve's conditional state, given Bob transmits symbol $k$, a Gaussian state with a $k$-independent von Neumann entropy.  It also makes her unconditional state identical to what prevails when Bob uses BPSK.  Consequently, applying the $\chi^{\rm UB}_{\rm EB}$ derivation from Ref.~\cite{Zhuang2016}'s Appendix~C to FL-QKD with KPSK requires only that in the final $\chi^{\rm UB}_{EB}$ formula---Eq.~(C56) of that appendix---the maximum allowable value for $\chi^{\rm UB}_{\rm EB}$ be increased from $R\log_2(2) = R$ to $R\log_2(K)$.  

\begin{figure}
\subfigure[]{
\includegraphics[width=0.22\textwidth]{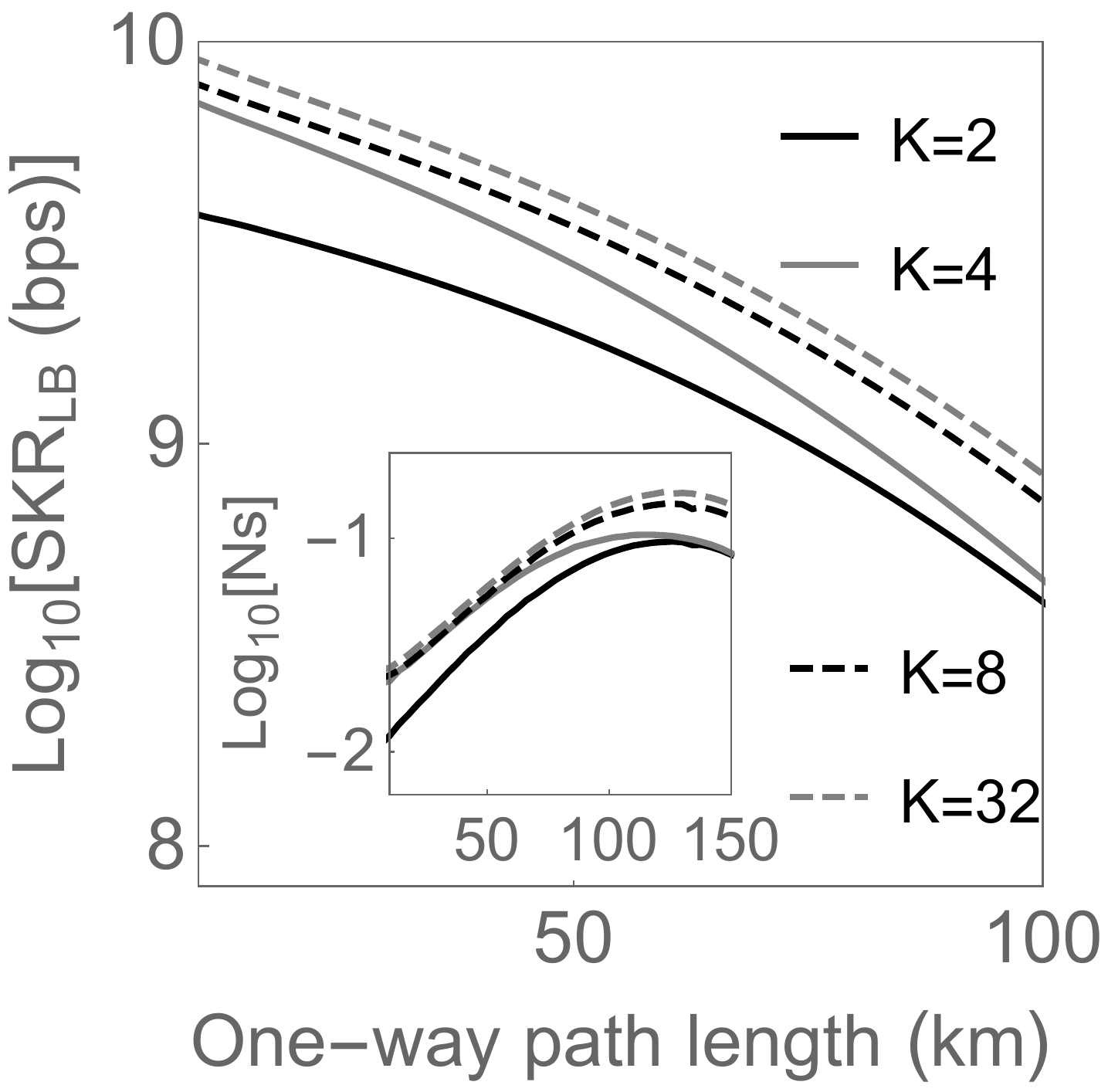}
\label{rate}}
\subfigure[]{
\includegraphics[width=0.22\textwidth]{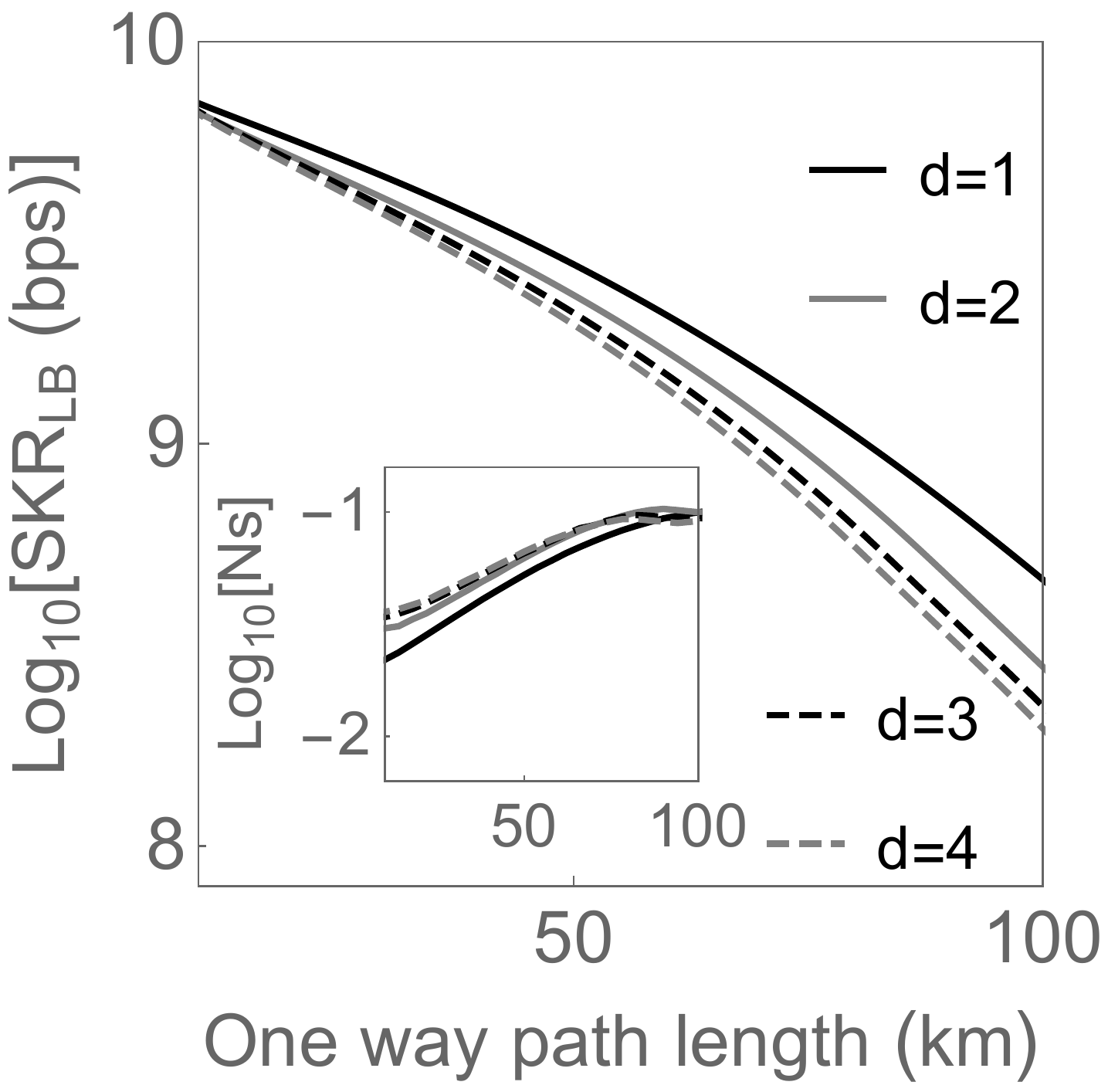}
\label{rate_QAM}}
\caption{SKR lower bounds for FL-QKD versus one-way path length $L$, with insets showing the optimized brightness $N_S$ of Alice's transmission to Bob.  The assumed parameter values are given in the text.  (a) ${\rm SKR}_{\rm LB}$ with KPSK for $1\le \log_2(K)\le 5$.  (b) ${\rm SKR}_{\rm LB}$ with $2d\times 2d$ square-lattice QAM for $1\le d\le 4$.}
\end{figure}

Figure~\ref{rate} plots KPSK's ${\rm SKR}_{\rm LB}$ versus one-way path length $L$ for $1\le \log_2{K} \le 5$ when, for each $L$ value, Alice's source brightness $N_S$ is chosen to maximize ${\rm SKR}_{\rm LB}$ and the other system parameters as given earlier in this section, i.e., they are the same as those employed in Ref.~\cite{Zhuang2016}.  We see that at 50\,km path length going from BSPK to 32PSK increases the SKR from 2\,Gbit/s to 4.5\,Gbit/s.  The inset in Fig.~\ref{rate} plots the optimized $N_S$ versus $L$.  As required to defeat Eve's passive eavesdropping attack on BPSK~\cite{Zhuang2016,Zhang2018,Shapiro2009}, we see that $N_S 
\ll 1$ prevails at all distances shown for that case.  Somewhat higher brightnesses---but still satisfying $N_S < 1$ at all distances---are optimum as $K$ increases, because Eve's decoding a higher-order KPSK requires her to have a higher-quality phase reference, something that is still inaccessible to her at those $N_S$ values.  As an interesting side note, we point out that the convergence with increasing $L$ of Fig.~\ref{rate}'s ${\rm SKR}_{\rm LB}$ 4PSK curve to its BPSK curve---a behavior that can be shown analytically---is due to the resulting in decrease Alice's signal-to-noise ratio and the structure of those two signal sets. 

\subsection{FL-QKD Performance with QAM Encoding\label{QAM}}

In FL-QKD with $2d\times 2d$ square-lattice QAM, Bob first selects a symbol from $0\le k \le K_q -1 \equiv 4d^2-1$ in an equiprobable manner.  He then intensity and phase modulates the light remaining after his channel-monitor tap to encode that symbol so that, in a noise-free world, it would appear as $\bar{I}_k + i\bar{Q}_k$---the center of the $k$th gray-shaded region in Fig.~\ref{scheme_modulation_QAM}---at the output of Alice's dual-homodyne receiver~\cite{footnote2}.   

Our first task is to use the Gaussian approximation for Alice's $I$ and $Q$ values' conditional distribution to determine Alice's minimum error-probability decision rule.  Finding that decision rule, without further approximation, is made difficult by the symbol-dependent conditional variances and covariance of $I$ and $Q$.  In Appendix~\ref{AppQAM}, however, we show that for the parameter values of interest, it is reasonable to take $I$ and $Q$ to be statistically independent, given Bob's transmitted symbol is $k$, with mean values $\bar{I}_k$ and $\bar{Q}_k$, and equal symbol-independent variances, $\sigma^2$.  

With the preceding approximation for QAM's conditional measurement statistics, $p(\,I,Q\mid k\,)$, finding FL-QKD's minimum error-probability decision rule reduces to the one for classical fiber-optic communication with QAM:  decoding an equiprobable QAM symbol from its observation in additive white Gaussian noise.  The minimum error-probability decision rule for that problem is minimum-distance decoding, i.e., the decision region, $\mathcal{D}_k$, for symbol $k$ is
\begin{equation}
\mathcal{D}_k = \{\,I+iQ : \arg\min_{k'} |(I-\bar{I}_{k'}) + i(Q-\bar{Q}_{k'}|\,\},
\end{equation}
as shown in Fig.~\ref{scheme_modulation_QAM}.   

So, to evaluate Alice and Bob's Shannon-information rate, we use QAM's $p(\,I,Q\mid k\,)$ and its $\mathcal{D}_{\tilde{k}}$ to calculate $\Pr(\,\tilde{k}\mid k\,)$ from Eq.~(\ref{transmatrix}) for $0\le k,k'\le K_q-1$.  The desired Shannon-information rate is then found from
\begin{eqnarray}
\lefteqn{I_{AB}=} \nonumber \\[.05in]
&&\!\!R\!\left\{\sum_{k=0}^{K_q-1}\sum_{\tilde{k}=0}^{K_q-1}\frac{\Pr(\,\tilde{k}\mid k\,)}{K_q}
\log_2\!\left[\frac{K_q\Pr(\,\tilde{k}\mid k\,)}{\sum_{k'=0}^{K_q-1}\Pr(\,\tilde{k}\mid k'\,)}\right]\right\}\!\!.
\label{IAB_QAM}
\end{eqnarray}

Now, to complete our goal of finding FL-QKD's ${\rm SKR}_{\rm LB}$ for operation with $2d\times 2d$ square-lattice QAM when Eve mounts an SPDC light-injection attack, we need to get $\chi^{\rm UB}_{EB}$ for that attack.  Eve's Holevo-information rate upper bound can be obtained in manner similar to the case of BPSK. Indeed, the derivation in Appendix~C of Ref.~\cite{Zhuang2016} is directly applicable with only minor changes.  This applicability is due to Eve's conditional state, given Bob transmits his $k$th symbol, still being Gaussian, and her unconditional state still having a Wigner covariance matrix that is diagonal.  Hence, Ref.~\cite{Zhuang2016}'s Appendix~C provides the upper bound we are seeking if we:  (1) take account of the $k$-dependent nature of Eve's conditional state in evaluating the average of her conditional-states' von Neumann entropies; (2) bound her unconditional state's von Neumann entropy by the von Neumann entropy of a thermal state with the same Wigner covariance matrix; and (3) use $R\log_2(K_q)$, instead of $R$, as the upper limit of her Holevo-information rate.  (See Appendix~\ref{AppQAM} for the details.)

Figure~\ref{rate_QAM} plots QAM's ${\rm SKR}_{\rm LB}$ versus one-way path length $L$ for $1\le d \le 4$ when, for each $L$ value, Alice's source brightness $N_S$ is chosen to maximize ${\rm SKR}_{\rm LB}$ and the other system parameters are the same as those employed in Ref.~\cite{Zhuang2016} and for KPSK.  The inset in this figure shows the optimized $N_S$ value versus $L$; as expected, low-brightness operation is maintained to ward off Eve's passive-eavesdropping attack.  What may not be expected for QAM, however, is the following behavior.
Unlike what we saw for KPSK---where increasing $K$ led to increasing SKR, albeit with diminishing returns, for $1\le \log_2(K) \le 5$---the best QAM performance, for $1\le d \le 4$, occurs when $d=1$.  But $d=1$ square-lattice QAM is merely quadrature phase-shift keying (QPSK = 4PSK) rotated by $\pi/4$\,rad, so we conclude that QAM, at least in its square-lattice form, offers no benefit SKR benefit to FL-QKD~\cite{footnote3}.  

\section{Discussion\label{Discussion}}
We have shown that 32PSK can increase FL-QKD's SKR on a 50-km-long fiber channel from 2.0\,Gbit/s to 4.5\,Gbit/s, but that square-lattice QAM offers no SKR improvement beyond its 4-ary case, which is equivalent to 4PSK.  Therefore, the first thing to discuss is the reason for this behavior, which contrasts sharply with QAM's ability to provide substantial capacity increases in classical fiber-optic communication by virtue of its higher spectral efficiency.  

It is easy to see why FL-QKD with KPSK suffers diminishing returns with increasing $K$.  Because $N_S < 1$ is maintained to ensure security against Eve's passive-eavesdropping attack, the $\{\bar{I}_k+i\bar{Q}_k\}$ become more tightly packed around a circle of limited radius in the $I,Q$ place with increasing $K$.  Thus, because the one-standard-deviation noise regions about these points do \emph{not} change with $K$, increasing $K$ makes it harder for Alice to reliably decode Bob's transmitted symbols, hence limiting Alice and Bob's SKR gain with increasing $K$.  

Alice's transmitting at low brightness, so that Eve cannot obtain a suitable phase reference to decode Bob's KPSK, defeats passive eavesdropping.  For $2d\times 2d$ square-lattice QAM with $d >1$, however, the situation is different.  Now, Bob's symbols vary in both intensity and phase.  So, even without a suitable phase reference, Eve's passive-eavesdropping attack can provide some intensity information about Bob's symbols.  Moreover, as is the case for KPSK, Alice faces increasing difficulty in discriminating between Bob's different QAM symbols with increasing $d$, because those symbols lie within a limited-radius circle in the $I,Q$ plane, and they are each surrounded by fixed-radius one-standard-deviation noise regions.  The result is that $d=1$ is the best of the $2d\times 2d$ square-lattice QAM constellations insofar as FL-QKD's SKR is concerned.  

In conclusion, FL-QKD---whether with its original BPSK encoding or with its high-order KPSK encoding---currently offers something that no other QKD protocol does:  Gbit/s SKRs over metropolitan-area distances with available technology and without the space-division or wavelength-division multiplexing.  Hence FL-QKD could make OTP encryption of high-data-rate traffic possible over such distances.  Two issues that remain to be addressed before widespread use of FL-QKD might occur are as follows.  

The first issue arises because FL-QKD is an interferometric protocol, which implies that proper functioning of Alice's dual-homodyne receiver requires that the roundtrip Alice-to-Bob-to-Alice fiber link be stabilized in time delay to < 1 ps and in phase to < 0.2 Rad for BPSK and even more finely for KPSK.  It turns out, however, that the BPSK-level challenge has been overcome by MIT Lincoln Laboratory, which has recently reported success in stabilizing the 86-km-roundtrip fiber link between its Lexington  Massachusetts location and the Cambridge Massachusetts MIT campus~\cite{Grein2017}.  Performing a field-test of FL-QKD on such a stabilized, deployed-fiber channel is the next experimental step that ought to be taken in FL-QKD's development.

The second issue to be addressed concerns FL-QKD's existing security proof's being limited to frequency-domain collective attacks in the asymptotic domain, as opposed, e.g., to decoy-state BB84's coherent-attack security proof with a finite-key correction~\cite{Lucamarini2013}.   Toward this end, a recent theoretical study~\cite{ZhuangZhu2017} elaborates the use of limited entanglement-assisted channel capacity to prove that Gaussian attacks are the optimum for a broad class of two-way QKD protocols.  We have used that result to establish a framework that could provide the desired coherent-attack security proof for FL-QKD~\cite{ZhuangZhang2017}.  Completing that security proof is the essential next step in FL-QKD's theoretical development.  

\begin{acknowledgements}
This research was supported by the Air Force Office of Scientific Research (AFOSR) MURI program under Grant No. FA9550-14-1-0052, and by the Office of Naval Research (ONR) under Contract No. N00014-16-C-2069. QZ also acknowledges support from the Claude E. Shannon Research Assistantship.
\end{acknowledgements}

\appendix

\section{Details for KPSK Encoding\label{AppPSK}}
\label{IAB_marray}
In this appendix we shall supply details for FL-QKD with KPSK encoding that were omitted from Sec.~\ref{KPSK}, viz., we will derive the means, variances, and covariance of $I$ and $Q$ conditioned on Bob's having transmitted the $k$th symbol from his KPSK alphabet.  Then, we will verify the rotational invariance claimed in Sec.~\ref{KPSK}, by proving that $\tilde{I}_k$ and $\tilde{Q}_k$ from Eqs.~(\ref{tildeI}) and (\ref{tildeQ}) are statistically independent, given $k$, and we will find their $k$-independent conditional variances, $\sigma^2_{\tilde{I}} > \sigma^2_{\tilde{Q}}$.    

Our work in this appendix will draw heavily upon the BPSK theory for FL-QKD that was established in Ref.~\cite{Zhuang2016}.  In that paper's Appendix~A, it was shown that the returned and LO light entering Alice's receiver for a particular symbol transmission comprise a collection of $M = TW$ iid mode pairs with photon annihilation operators $\{\,\hat{a}_{B_m}': 1\le m \le M\,\}$ and $\{\,\hat{a}_{R_m}' : 1\le m \le M\,\}$, respectively.  Under Eve's SPDC light-injection attack, and assuming Bob has transmitted his $k$th symbol, the results from Ref.~\cite{Zhuang2016}'s Appendix~A---generalized to account for KPSK signalling---show that the $\hat{a}_{B_m}'$ and $\hat{a}_{R_m}'$ modes are in a zero-mean, jointly-Gaussian state that is completely characterized by its non-zero second moments:
\begin{align}
\braket{\hat{a}_{B_m}^{\prime\dagger} \hat{a}_{B_m}'}&=
\kappa_S[G_B (1-\kappa_B)  \kappa_SN_S+N_B] \equiv n_B, \\[.05in]
\braket{\hat{a}_{R_m}^{\prime\dagger} \hat{a}_{R_m}'}&=  N_{\rm LO}, \\[.05in]
\braket{\hat{a}_{B_m}^{\prime\dagger} \hat{a}_{R_m}'} & = e^{-2\pi ik/K}c_{RB},
\end{align}
where 
\be 
c_{RB}\equiv \kappa_S[G_B (1-\kappa_B) (1-f_E) N_SN_{\rm LO}n/(n+1)]^{1/2}.
\ee

Alice 50-50 beam splits her returned and LO modes to provide the following inputs for the $I$ and $Q$ channels of her dual-homodyne receiver: 
\begin{eqnarray}
\hat{a}_{B_m}^{\prime (I)}&=&(\hat{a}_{B_m}^{\prime }+\hat{a}_{V_{B_m}})/\sqrt{2},\\[.05in]
\hat{a}_{B_m}^{\prime (Q)}&=&(\hat{a}_{B_m}^{\prime }-\hat{a}_{V_{B_m}})/\sqrt{2},\\[.05in]
\hat{a}_{R_m}^{\prime (I)}&=&(\hat{a}_{R_m}^{\prime }+\hat{a}_{V_{R_m}})/\sqrt{2},\\[.05in]
\hat{a}_{R_m}^{\prime (Q)}&=&(\hat{a}_{R_m}^{\prime }-\hat{a}_{V_{R_m}})/\sqrt{2},
\end{eqnarray}
where the $\{\hat{a}_{V_{B_m}}, \hat{a}_{V_{R_m}}\}$ are in their vacuum states.  The $I$ and $Q$ outputs of Alice's dual-homodyne receiver are then the results of the quantum measurements
\begin{align}
\hat{I} &= \sum_{m=1}^M(\hat{a}_{+m}^{\prime (I)\dagger}\hat{a}_{+m}^{\prime (I)} - 
\hat{a}_{-m}^{\prime (I)\dagger}\hat{a}_{-m}^{\prime (I)}), \\[.05in]
\hat{Q} &= \sum_{m=1}^M(\hat{a}_{+m}^{\prime (Q)\dagger}\hat{a}_{+m}^{\prime (Q)} - 
\hat{a}_{-m}^{\prime (Q)\dagger}\hat{a}_{-m}^{\prime (Q)}),
\end{align}
where 
\begin{align}
\hat{a}_{\pm m}^{\prime (I)} &\equiv \sqrt{\eta}\,(\hat{a}_{B_m}^{\prime (I)} \pm \hat{a}_{R_m}^{\prime (I)})/\sqrt{2} +\sqrt{1-\eta}\,\hat{v}_{\pm m}^{(I)},\\[.05in]
\hat{a}_{\pm m}^{\prime (Q)} &\equiv \sqrt{\eta}\,(\hat{a}_{B_m}^{\prime (Q)} \pm i\hat{a}_{R_m}^{\prime (Q)})/\sqrt{2} +\sqrt{1-\eta}\,\hat{v}_{\pm m}^{(Q)},
\end{align}
with $\eta$ being her receiver's homodyne efficiency, and the $\{\hat{v}_{\pm m}^{(I)},\hat{v}_{\pm m}^{(Q)}\}$ modes being in their vacuum states.

Straightforward calculations now yield the following expressions for the conditional means, variances, and covariance of $I$ and $Q$, given that Bob's transmitted symbol was $k$:
\begin{eqnarray}
\bar{I}_k &=& M\eta\cos(2\pi k/K)c_{RB},\\[.05in]
\bar{Q}_k &=& M\eta\sin(2\pi k/K)c_{RB},\\[.05in]
\sigma^2_{I_k}&=& M\eta[\eta \cos(4\pi k/K)c_{RB}^2+n_B\nonumber \\[.05in]
&&+\,N_{\rm LO}(1+\eta n_B)]/2, \\[.05in]
\sigma_{Q_k}^2 &=& M\eta[-\eta\cos(4\pi k/K)c_{RB}^2 + n_B \nonumber \\[.05in]
&& +\,N_{\rm LO}(1+\eta n_B)]/2,\\[.05in]
\sigma_{I_kQ_k} &=& M\eta^2\sin(4\pi k/K)c_{RB}^2/2.
\end{eqnarray}
Finally, we can prove our rotational invariance claim for the conditional statistics of $I$ and $Q$.  Using Eqs.~(\ref{tildeI}) and (\ref{tildeQ}), together with the conditional moments we have just obtained, gives us the desired result:  $p(\,\tilde{I}_k,\tilde{Q}_k \mid k\,)$ is a Gaussian distribution that is completely characterized by the following moments,
\begin{align}
&\langle\tilde{I}_k\rangle = M\eta c_{RB},\\[.05in]
&\langle \tilde{Q}_k\rangle = 0,\\[.05in]
&\sigma^2_{\tilde{I}_k} = M\eta[\eta c^2_{RB} + n_B + N_{\rm LO}(1+\eta n_B)]/2,\\[.05in]
&\sigma^2_{\tilde{Q}_k} = M[-\eta c^2_{RB} + n_B + N_{\rm LO}(1+\eta n_B)]/2, \\[.05in]
&\sigma_{\tilde{I}_k\tilde{Q}_k} = 0.
\end{align}
Here, the Gaussian nature of $p(\,\tilde{I}_k,\tilde{Q}_k \mid k\,)$ follows from $p(\,I,Q \mid k\,)$'s being Gaussian, and the statistical independence of $\tilde{I}_k$ and $\tilde{Q}_k$ given $k$ was sent then follows from their being uncorrelated ($\sigma_{\tilde{I}_k\tilde{Q}_k} = 0$).  We also see that $\tilde{I}_k$ and $\tilde{Q}_k$ have $k$-independent variances, $\sigma^2_{\tilde{I}} > \sigma^2_{\tilde{Q}}$, given $k$ was sent.

\section{Details for QAM Encoding \label{AppQAM}}
Here we shall supply details for FL-QKD with QAM encoding that were omitted from Sec.~\ref{QAM}, i.e., the conditional means, variances, and covariance of $I$ and $Q$, where we again rely on results from Ref.~\cite{Zhuang2016}'s Appendix A.  

Suppose that Bob encodes his $k$th symbol on the light remaining after his channel monitor's tap by imposing a transmissivity $0< \kappa_q \le 1$ attenuation and a $0\le \theta_q < 2\pi$ phase shift that are chosen in accord with where the $k$th symbol appears in his $2d\times 2d$ square-lattice QAM constellation.  Conditioned on that symbol being sent, and assuming that Eve has mounted an SPDC light-injection attack, the returned and LO light that enters Alice's receiver are again comprised of $M$ iid $\{\hat{a}_{B_m}',\hat{a}_{R_m}'\}$ mode pairs that are each in a zero-mean, jointly-Gaussian state that is completely characterized by its non-zero second moments:
\begin{eqnarray}
\braket{\hat{a}_{B_m}^{\prime\dagger} \hat{a}_{B_m}'}&=&
\kappa_S[G_B (1-\kappa_B)\kappa_q  \kappa_SN_S+N_B] \equiv n_{B_q} ,
\nonumber\\
\braket{\hat{a}_{R_m}^{\prime\dagger} \hat{a}_{R_m}'}&=&  N_{\rm LO},
\nonumber\\
 \braket{\hat{a}_{B_m}^{\prime\dagger} \hat{a}_{R_m}' }&=& \sqrt{\kappa_q}\,e^{-i\theta_q}c_{RB}.
\end{eqnarray}

It is now a relatively simple matter to show that 
\begin{align}
\bar{I}_k &=M\eta\sqrt{\kappa_q}\,\cos(\theta_q)c_{RB},\\[.05in]
\bar{Q}_k &=M\eta\sqrt{\kappa_q}\,\sin(\theta_q)c_{RB},
\end{align}
are the conditional-mean values, and 
\begin{eqnarray}
\sigma_{I_k}^2 &=& M\eta[\eta\kappa_qc_{RB}^2\cos(2\theta_q)+ n_{B_q} \nonumber \\[.05in]
&&+\,N_{\rm LO}(1+\eta n_{B_q})]/2, \label{varIk} \\[.05in]
\sigma_{Q_k}^2 &=& M\eta[-\eta\kappa_qc_{RB}^2\cos(2\theta_q)+ n_{B_q} \nonumber \\[.05in]
&& +\,N_{\rm LO}(1+\eta n_{B_q})]/2,\label{varQk}\\[.05in]
\sigma_{I_kQ_k}^2 &=&M\eta^2\kappa_qc_{RB}^2\sin\left(2\theta_q\right)/2,\label{covIkQk}
\end{eqnarray}
are the conditional variances and covariance of Alice's $I$ and $Q$ values when Bob transmits his $k$th symbol.  

To simplify finding Alice's minimum error-probability decision regions---and hence the calculation of Alice and Bob's Shannon-information rate---we note that the parameter values assumed in Sec.~\ref{SKRs} imply that
\begin{align}
N_{\rm LO} &\gg \kappa_qc_{RB}^2 \sim \kappa_q\kappa_S^2G_BN_SN_{\rm LO} \nonumber \\[.05in] 
&> n_{B_q} \sim \kappa_q\kappa_SN_B \gg 1.
\end{align}
Using these relations simplifies Eqs.~(\ref{varIk})--(\ref{covIkQk}) to
$\sigma_{I_k}^2  \approx M\eta^2N_{\rm LO}n_{B_q}/2, \sigma_{Q_k}^2  \approx  M\eta^2N_{\rm LO}n_{B_q}/2,$ and 
$\sigma_{I_kQ_k} \approx 0$, thus justifying the white Gaussian noise assumption made in Sec.~\ref{QAM}.


\begin{thebibliography}{99}
\bibitem{footnote2}The corner points in the Fig.~\ref{scheme_modulation_QAM} signal constellation are produced with just phase shifts.  The remaining points require attenuation, with those that do not lie on the $I$ axis requiring phase shifts as well.
\bibitem{footnote3}This point is further underscored by our not having proved Eve's SPDC light-injection attack
realizes her optimum frequency-domain collective attack on $2d\times 2d$ square-lattice QAM for $d > 1$.
\end{thebibliography}
\end{document}